\documentclass[3p,times,twocolumn]{elsarticle}

\usepackage{ecrc}
\usepackage{epsfig}
\usepackage{subfigure}

\volume{00}

\firstpage{1}

\journalname{Nuclear Physics B Proceedings Supplement}

\runauth{Z. Was}

\jid{nuphbp}

\jnltitlelogo{Nuclear Physics B Proceedings Supplement}

\usepackage{amssymb}

\usepackage[figuresright]{rotating}

\begin{document}

\begin{frontmatter}



\dochead{}

\title{New hadronic currents in TAUOLA:
for confrontation with the experimental data}


\author{ Z. W\c{a}s}

\address{Institute of Nuclear Physics, Polish Academy of Sciences,\\
         ul. Radzikowskiego 152, 31-342 Cracow, Poland \\ and \\
CERN, 1211 Geneva, Switzerland}

\begin{abstract}
The status of implementation of new hadronic currents into the Monte Carlo system for simulation of $\tau$-lepton production and decay in
high-energy accelerator experiments is reviewed.
Since the $\tau$-lepton conference in 2010 substantial  progress was achieved:
 (i) For the {\tt TAUOLA} Monte Carlo generator of $\tau$-lepton decays,
automated and simultaneous use of many versions of
form factors for the calculation of optional weights  for fits was developed
and checked to work in the Belle and BaBar software environment.
Alternative parameterizations of hadronic currents based on the Resonance Chiral approach are available now. 
This was achieved for more than 88\% of the total $\tau$ hadronic width.
(ii)
the  {\tt TAUOLA universal interface} based on {\tt HepMC} (the {\tt C++}
event record) is available. This is the case for C++ users of {\tt PHOTOS} Monte Carlo
for radiative corrections in decays, as well. An algorithm for weighted
events to explore spin effects in analysis of hard processes was prepared.
(iii) Kernels featuring a complete first-order matrix element
are available now for  {\tt PHOTOS} users interested in  decays of $Z$ and $W$ bosons. New
tests with different options of matrix elements for those and for
$K_{l3}$ decays are available as well.

Presented results illustrate  the status of the projects performed in
collaboration with  Zofia Czyczula, Nadia Davidson,
Tomasz Przedzi\'nski, Olga Shekhovtsova,
El\.zbieta Richter-W\c{}as, Pablo Roig, Qingjun Xu and others.

\vskip 3 mm
\centerline{preprint \hskip 1 cm {\bf  CERN PH-TH/2012-012}, \hskip 1 cm January 2012 \hskip 7.5 cm}
\end{abstract}

\begin{keyword}

lepton tau   \sep  Resonance Chiral Theory  \sep Bremmstrahlung
 \sep Monte Carlo \sep TAUOLA \sep PHOTOS
\end{keyword}

\end{frontmatter}

\section{Introduction}

The {\tt TAUOLA} package
\cite{Jadach:1990mz,Jezabek:1991qp,Jadach:1993hs,Golonka:2003xt} for simulation
of $\tau$-lepton decays and
{\tt PHOTOS} \cite{Barberio:1990ms,Barberio:1994qi,Golonka:2005pn} for simulation of QED radiative corrections
in decays, are computing
projects with a rather long history. Written and maintained by
well-defined (main) authors, they nonetheless migrated into a wide range
of applications where they became ingredients of
complicated simulation chains. As a consequence, a large number of
different versions are presently in use. Those modifications, especially in case of
{\tt TAUOLA}, are   valuable from the physics point of view, even though they
 often did not find the place in the distributed versions of
the program.
From the algorithmic point of view, versions may
differ only in  details, but they incorporate many specific results from distinct
$\tau$-lepton measurements or phenomenological projects.
Such versions were mainly maintained (and will remain so)
by the experiments taking precision data on $\tau$ leptons.
Interesting from the physics point of view changes are still
developed in {\tt FORTRAN}.
That is why, for convenience of such partners, part of the
{\tt TAUOLA} should remain in {\tt FORTRAN} for a few forthcoming years.

Many new applications were developed in C++,  often requiring
a program interface to other packages  (e.g., generating events for LHC, LC,
Belle or BaBar physics processes).
Fortunately, co-existence of {\tt FORTRAN} with C++ is not a problem, at least
not from
the software point of view.

The program structure,
 was presented during
$\tau$ conferences,
and we will not repeat it here.
This time, let us concentrate on new hadronic currents based on the Resonance Chiral approach.
We will also report on prepared
techniques useful  for fits.
Analyses of high precision,
high-statistics  data from Belle and BaBar are expected to progress from these
solutions. Other aspects of the project such as interfaces
for applications based on {\tt HepMC} \cite{Dobbs:2001ck} event record
or new tests and weighting  algorithms for spin effects
in production processes will be  mentioned as well.

Our presentation is organized as follows:
Section 2  is devoted to the discussion  of optional
weights in {\tt TAUOLA} and their use for fits to experimental data.
Status of implementation of new
currents for hadronic decays which can be confronted with (tuned to)
data using such weights enabling simultaneous control of all
experimental effects will be  mentioned, but results of this work are covered
in another talk of the conference.
In section 3 we concentrate on   {\tt PHOTOS} Monte Carlo for
radiative corrections in decays.
Section 4 is devoted to new interfaces of {\tt TAUOLA} and {\tt PHOTOS} based
on  {\tt HepMC} and written in C++. Work on interface to
genuine weak corrections, transverse spin effects and new tests
and implementation
bremsstrahlung kernels will be presented as well.
Short section 5, is devoted to {\tt MC-TESTER}; the program designed for
semi-automatic comparisons of simulation samples originating from
different programs and heavily used in our projects.

Because of the limited space of the contribution,
and sizeable amount of
results, some of them will not be given in the
proceedings. They find their place in
publications, prepared with coauthors listed in the Abstract.
For these works,  the present paper may serve as a summary.

\section{ Approach of Resonance Chiral lagrangians and   {\tt TAUOLA} Monte Carlo}

\begin{figure}
\subfigure{
\includegraphics[scale=.330]{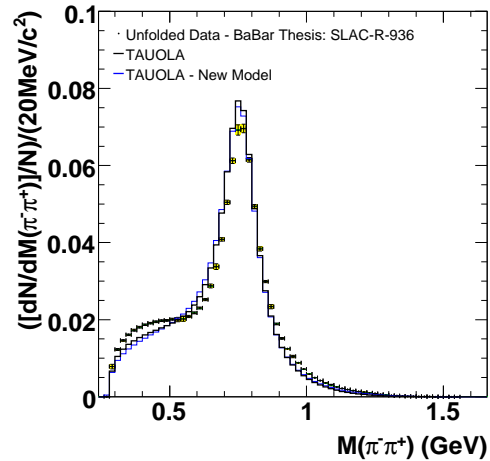}}
\subfigure{
\includegraphics[scale=.330]{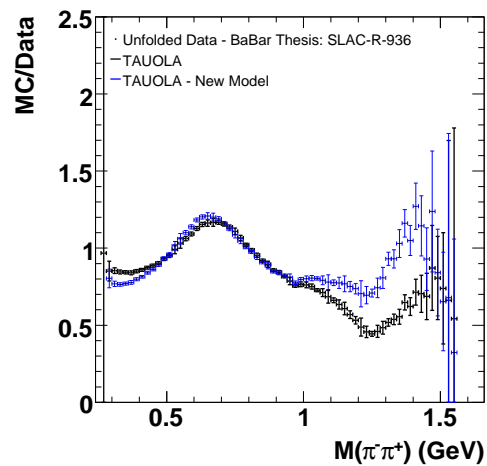}}
\caption{Invariant mass distribution of the $\pi^+\pi^-$ pair in $\tau \to \pi^+\pi^- \pi^-\nu_\tau$ decay.
Lighter grey histogram is from our new model, darker grey is from default
parametrization of {\tt TAUOLA cleo}.  The unfolded BaBar data are
taken from Ref.~\cite{Nugent:2009zz} \label{Test}.
The plot on top presents
distribution in invariant mass of $\pi^+\pi^-$ pair, and the one on bottom
 ratios
between Monte Carlo results and data (courtesy of Ian Nugent).}
\end{figure}

In another talk \cite{Roig:2011iv} of the conference, an approach,
based on Resonance Chiral Lagrangian, for calculations of hadronic currents
 to be used in {\tt TAUOLA} was
described. That is why, we do not need to repeat it here. In \cite{RChL}
implementation of those currents is documented in a great detail.
Technical tests are available with this reference as well. Let us limit
ourselves  to one example, Fig.~\ref{Test}.

Physics of $\tau$ lepton decays requires sophisticated strategies for the
confrontation of phenomenological models with experimental data. On one hand,
high-statistics experimental samples are collected, and the obtained precision is
high, on the other hand, there is a significant cross-contamination between distinct
$\tau$ decay channels. Starting from  a certain precision  level all channels
need to be analyzed simultaneously. Change of parameterization for one channel
contributing  to the background to another one may be important for the fit of
its currents. This situation leads to a complex configuration where a multitude of parameters (and models)
needs to be simultaneously confronted with a multitude of observables.
One has to keep in mind that the models used to obtain distributions in
 the fits may require refinements or even substantial rebuilds as a consequence
of comparison with the data. The topic was covered in detail in the $\tau$ Section of Ref.~\cite{Actis:2010gg}.

From the statistical point of view it is best to resolve such a system in one
automated step using, for example, a method
such as described in \cite{tmva,Hocker:2007ht}.
This can be of course very dangerous from
the point of view of systematic error control. But we will not elaborate on this
point any further. From the technical side
 it is necessary to calculate for each generated event (separately for each
present in it decay of $\tau^+$ and/or $\tau^-$) alternative weights; the ratios
of the matrix element squared obtained with new currents,
and the one actually used in generation. Then, the vector of weights can be obtained
and used in fits.
We have checked that such a solution not only can be easily installed into
{\tt TAUOLA} as a stand-alone generator, but it can also be incorporated into
the simulation frameworks of Belle and BaBar collaborations.
The weights can be calculated after the simulation of detector response is
completed. Only then choice of parameters for the hadronic currents has to be
performed and the fits completed.

We take into account convenience for necessary  software upgrades
in experiments. Instead of a
completely new system, only
a dedicated patch is prepared.

\section{{\tt PHOTOS} Monte Carlo for bremsstrahlung and its systematic uncertainties}
\def\CCol{{\tt SANC}}
Thanks to exponentiation properties and factorization, the bulk of the final state
QED bremsstrahlung can be described in a universal way.
However, the
kinematic configurations caused by QED bremsstrahlung are affecting
in an  important way
signal/background separation. It may affect selection criteria and background
contaminations in quite complex and unexpected ways.
In many applications, not only in $\tau$ decays,
such bremsstrahlung corrections are
generated with the help of the  {\tt PHOTOS} Monte Carlo. That is why it is of importance to
review the precision of this program as documented in
Refs.~\cite{Barberio:1990ms,Barberio:1994qi,Golonka:2005pn}.
For the C++ applications, the version of the program is available now.
It is documented in Ref.~\cite{Davidson:2010ew}.

In C++ applications, the complete first-order matrix elements  for the
 two-body decays of the $Z$ \cite{Golonka:2006tw} and $W$ \cite{Nanava:2009vg}
decays into a lepton pair are now available.
Kernels with complete matrix elements, for the decays of
scalar $B$ mesons  into a pair of scalars  \cite{Nanava:2006vv} are
available for the C++ users as well.
For  $K \to l \nu \pi$
and  for $\gamma^* \to \pi^+\pi^-$ decays \cite{Nanava:2009vg,Xu:2012px}
matrix element based kernels are still available for tests only.
Properly  oriented reference frames are needed in those cases.
It will be rather easy to
integrate those NLO kernels  into the main version of the program,
 because of better control of the
decay particle rest frame than in the {\tt FORTRAN} interface.

In all of these cases the universal kernel of {\tt PHOTOS} is replaced with the
one matching an exact first-order matrix element. In this way terms necessary
for the NLO/NLL precision 
level are implemented\footnote{Note that here the LL (NLL) denotes
 collinear logarithms (or in case of differential
predictions terms integrating into such logarithms).
 The logarithms of soft singularities are taken into
account to all orders. This is resulting from mechanisms of exclusive
exponentiation \cite{Jadach:2000ir} of QED.
The algorithm used in {\tt PHOTOS} Monte Carlo is compatible with exclusive
exponentiation. Note that our
 LL/NLL precision level would even read  as  respectively   NLL/NNNLL
 level in some naming conventions of QCD.
}.
A discussion relevant for control of program systematic uncertainty in $\tau \to \pi \nu$ decay can be found in
Ref.~\cite{Guo:2010ny}.

The algorithm covers the full multiphoton
phase-space and becomes  exact in the soft limit.
This is rather unusual for  NLL compatible algorithms. One should not forget
that {\tt PHOTOS} generates weight-one events, and does not exploit any
phase space ordering. There is a full phase space overlap between the one
where a hard
matrix element is used and the one for iterated photon emission.
All interference effects (between consecutive emissions and emissions from
distinct charged lines) are implemented with the help of internal weights.

The results of all tests of {\tt PHOTOS} with a NLO kernel confirm
 sub-permille precision level.
This is very encouraging, and points to the possible extension of the
approach outside of  QED (scalar QED). In particular, to the domain of
QCD or to QED when  phenomenological form factors for interactions
of photons need
 to be used. For that work  to be completed, spin amplitudes need
to be studied. Let us point to Ref.~\cite{vanHameren:2008dy}
as an example.

New tests of {\tt PHOTOS} are available from the web page \cite{Photos_tests}.
In those tests,
results from the second-order matrix element calculations
embedded in KKMC \cite{Jadach:1999vf} Monte Carlo are used in case of $Z$ decay.
For $W$ decays comparisons with electroweak calculations of
Refs.~\cite{Andonov:2008ga,Andonov:2004hi} are shown.

\section{  {\tt TAUOLA universal interface} and {\tt PHOTOS} interface in C++}

\begin{figure}
\begin{center}
\setlength{\unitlength}{0.5 mm}
\begin{picture}(35,80)
\put( -65,-45){\makebox(0,0)[lb]{\epsfig{file=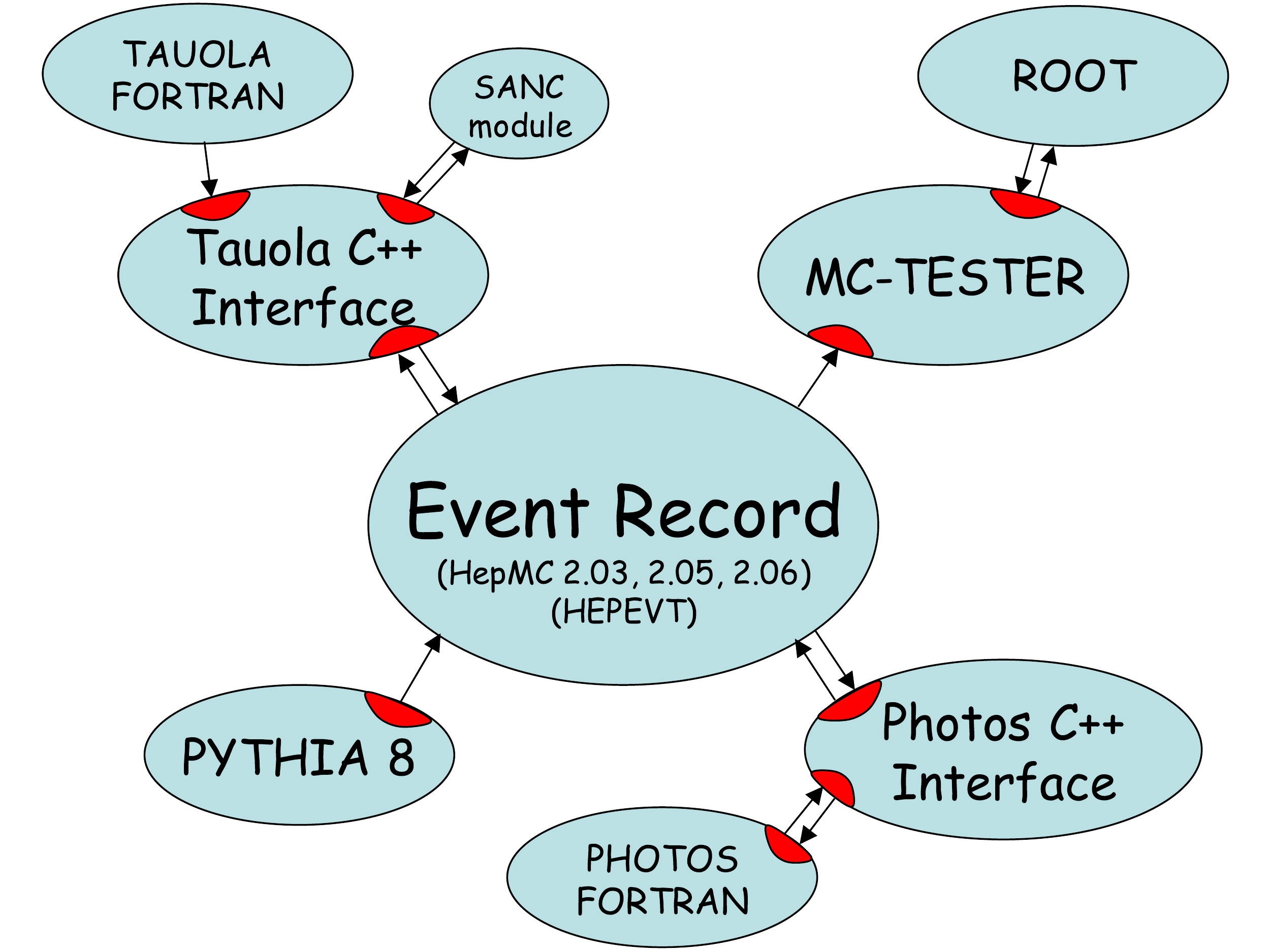,width=70mm,height=65mm}}}
\end{picture}
\end{center}
\vskip 1.5 cm
\caption{\small \it  Scheme of Monte Carlo simulation system with communication based on event record.
 } \label{Relations}
\end{figure}

In the development of packages such as {\tt TAUOLA} or {\tt PHOTOS}, questions
of tests and appropriate relations to users' applications are essential for
their
usefulness. In fact, user applications may be much larger in size and
human efforts than the programs discussed here.
Good example of such `user applications' are complete environments to simulate
physics process and control detector response at the same time.
Distributions of final state particles are not always of direct interest.
Often properties of intermediate states, such as a spin state of $\tau$-lepton,
coupling constants or masses of intermediate heavy particles are
of prime interest.
As a consequence, it is useful that such intermediate state properties are
under direct control of the experimental user and can be manipulated
to understand detector responses.
Our programs  worked well   with {\tt FORTRAN} applications  where {\tt HEPEVT} event record
is used.  For the  {\tt C++}  {\tt HepMC} \cite{Dobbs:2001ck} case,
interfaces were  rewritten, both for
{\tt TAUOLA} \cite{Davidson:2010rw} and for {\tt PHOTOS} \cite{Davidson:2010ew}.
The interfaces and as a consequence the programs themselves
were enriched; for  {\tt PHOTOS} new Matrix element kernels
are  available; for {\tt TAUOLA} interface,
 a complete (not  longitudinal only) spin correlations
 are available
for $Z/\gamma^*$ decay.
Electroweak corrections taken from
Refs.~\cite{Andonov:2008ga,Andonov:2004hi} are also used.
For the scheme
of programs communications see Fig.~\ref{Relations}.
In this spirit an algorithm \cite{Czyczula:2012ny} to study detector response to spin effects
in $Z, W$ and $H$ decays, was developed.
Such modular organization opens ways for further efficient algorithms to understand
detector systematics, but at the same time responsibility to control software
precision must be shared by the user.
For that purpose
automated   tests of
{\tt MC-TESTER} were prepared~\cite{Golonka:2002rz}.
New functionalities
were introduced into the testing package \cite{Davidson:2008ma}. In particular, it works now with the
{\tt HepMC} event record, the  standard of {\tt C++} programs and the
spectrum of available tests is enriched.

\section{{\tt MC-TESTER} its user defined tests and Grid libraries.}
Our work on {\tt MC-TESTER} reached maturity with  Ref.~\cite{Davidson:2008ma}.
As in the past, the program
main purpose remains benchmarking the decay part of different Monte Carlo chains.
Generated events have to be  stored in event records: be it of {\tt FORTRAN}
 or C++.  Default distributions consist of all
possible invariant masses which are automatically generated and stored for
each found decay channel of the particle under test.
Then, at the analysis step,  information from a pair of such runs may
 be compared and represented in the form of tables and plots.
At present, users macros can be easily installed, in particular all
demo distributions given in papers on C++ interfaces for
{\tt TAUOLA} \cite{Davidson:2010rw} and
{\tt PHOTOS} \cite{Davidson:2010ew} were obtained in that way.

Set-up's  for benchmarking the interfaces, such as interface
 between $\tau$-lepton
production and decay, including QED bremsstrahlung effects
are also prepared in that way.
The updated version of {\tt MC-TESTER} was found useful
for {\tt FORTRAN} ~\cite{Golonka:2005pn,Golonka:2006tw}
and for C++ \cite{Davidson:2010ew}  examples
 where spurious information (on soft photons)
was removed.

Finally, let us mention that the program is available
through the Grid Project LCG/Genser web page, see Ref.~\cite{Kirsanov:2008zz} for details.
 This is the case for {\tt TAUOLA} C++ and for
{\tt PHOTOS}  C++  as well.
The {\tt FORTRAN} predecessors have already been
available in this and for a longer time now.

\section{Summary and future possibilities}

Versions of the hadronic currents available for the {\tt TAUOLA} library
until now, are all based on old models and experimental data of 90's.
Implementation of   new currents, based on the Resonance Chiral Lagrangian approach
 is now prepared
and tested from the technical side. Methods for efficient confrontation
with the experimental data are prepared as well.
Once comparison with Belle and BaBar data successfully completed,
new parameterizations will be straightforward for use in
a broad spectrum of applications
in {\tt FORTRAN} and C++ environments.

The status of
associated projects: {\tt TAUOLA universal interface } and {\tt MC-TESTER}
was reviewed. Also
the high-precision version of  {\tt PHOTOS} for radiative corrections in
decays, was
presented. All these programs are available now for C++ applications
thanks to the {\tt HepMC} interfaces.

New results for  {\tt PHOTOS}   were mentioned.
For the leptonic $Z$ and $W$ decays  the complete next-to-leading collinear logarithms
effects can now be  simulated in C++ applications.
However, in  most cases these
effects are not important, leaving the standard  version
 sufficient.
 Thanks to this work  the path for  fits to the data of
electromagnetic form factors
 is opened, e.g. in the  case of $K_{l3}$ decays.

 The presentation of the {\tt TAUOLA} general-purpose interface
in {\tt C++} was given. It is more refined
than the {\tt FORTRAN} predecessor. Electroweak corrections can be used
in calculation of complete spin correlations in $Z/\gamma^*$ mediated
processes.  An algorithm for study of detector responses to spin effects in
$Z$, $W$ and $H$ decays was shown.

The present version of {\tt MC-TESTER} is stable now.
It works with {\tt HepMC}  of {\tt C++} and enables
user defined tests in experiments' software environments.  We used the tool
regularly all over  our projects.

\vskip 1 mm
\centerline{ \bf Acknowledgements}
\vskip 1 mm

Discussions
with  members of the Belle and BaBar collaborations
are  acknowledged. Exchange of e-mails and direct discussions
with  S. Banerjee, S. Eidelman, H. Hayashii, K. Inami, J. H. K\"uhn   and
M. Roney was a valuable  input to present and future steps in
projects development.

\providecommand{\href}[2]{#2}                                   
\end{document}